# Symmetric Trotterization in digital quantum simulation of quantum spin dynamics

Yeonghun Lee[1,*]

[1] Department of Electronics Engineering, Incheon National University, Incheon 22012, Republic of Korea

**Corresponding author**

[*] Email: y.lee@inu.ac.kr

**Abstract**

A higher-order Suzuki-Trotter decomposition or Trotterization can be exploited to mitigate the Trotter error in digital quantum simulation. This work revisits the second-order symmetric Trotterization in terms of the Trotter error, where quantum many-body spin dynamics of the transverse-field Ising model is simulated. While the work presents a pedagogical way to exploit a real quantum computer, the effectiveness of the symmetric Trotterization is evaluated in a prototype superconducting quantum device on IBM Quantum Experience. It turns out that the symmetric Trotterization does not provide higher accuracy than the first-order Trotterization in the testbed using the transverse-field Ising model. The result indicates that apart from the quantum errors, such as logical gate error and readout error, the use of a higher-order

Trotterization should be circumspect, and the Trotter error would play an insignificant role in particular applications in an early stage of realized noisy intermediate-scale quantum (NISQ) devices.

# 1. Introduction

Quantum computing has the potential to solve classically intractable problems with exponential speed-up [1], where programmable digital quantum simulation (DQS) allows us to explore the unitary evolution of any many-body Hamiltonian [2]. While fault-tolerant quantum computing remains the long-term goal, currently realized noisy intermediate-scale quantum (NISQ) devices [3] have demonstrated DQSs for a wide variety of applications—quantum chemistry [4–7], quantum many-body spin dynamics [4,8–13], and strongly correlated systems [14–16]—although they have not outperformed the classical counterparts due to significant quantum errors.

The Suzuki-Trotter decomposition [17,18] discretizes the unitary evolution and creates a sequence of quantum gates. In DQS, the Trotterization is a common approach not only to simulate unitary time evolution for quantum dynamics but also to find the ground state using the variational quantum eigensolver (VQE) [19,20] or quantum imaginary time evolution (QITE) [21]. On top of the low fidelity of quantum gate operations in NISQ devices, Trotterization incurs an inherent error, called Trotter error, where a higher-order Suzuki-Trotter decompostion [22,23] can be advantageous to reduce the Trotter error. Thus, a second-order Trotterization (symmetric Trotterization) is often used for error mitigation in DQS of physical systems [8,12,24,25].

The transverse-field Ising model (TFIM), one of the simplest quantum many-body systems, has been employed as a testbed for quantum simulation [8–10,26–31]. The spin operators can transform into fermionic creation and annihilation operators through spin-to-fermion mappings, such as the Jordan-Wigner transformation [32]. Despite its simple form, TFIM possesses interesting physics and plays a role in the field of quantum computing. For example, the fermionized Hamiltonian of TFIM with open boundary conditions resembles a one-dimensional p-wave superconductor and presents Majorana modes [33]. TFIM can be a platform for investigating quantum chaos and eigenstate thermalization [34]. Besides, the quantum annealing (QA) [35] and the quantum approximate optimization algorithm (QAOA) [36] can be interfaced with TFIM to prepare a ground state.

Quantum information technology has been evolving rapidly, and the study of Trotterization is timely. Spin many-body systems are of importance to not only condensed matter physics but also quantum simulation, where the direct correspondence between qubit and spin facilitates the study of Trotterization in DQS. This work investigates the effectiveness of the symmetric Trotterization in DQS for far-from-equilibrium dynamics of transverse-field Ising chains.

## 2. Transverse-field Ising model

A Hamiltonian of TFIM is composed of the interaction term and the transverse-field term:

$$\hat{H} = -J \sum_{j}^{N} \hat{\sigma}_j^z \hat{\sigma}_{j+1}^z - g \sum_{j}^{N} \hat{\sigma}_j^x, \quad (1)$$

where $\hat{\sigma}_j^z$ and $\hat{\sigma}_j^x$ are the Pauli operators for the $j$th spin, $N$ is the total number of spins, $J$ is the coupling strength, and $g$ is the transverse magnetic field. A quantum spin state at time $t$ is given by the unitary time evolution $|\Psi(t)\rangle = e^{-i\hat{H}t}|\Psi(0)\rangle$. In a first-order Trotterization, the time evolution operator can be rewritten in [10]

$$e^{-i\hat{H}t} = \left( \prod_{j \text{ even}} \hat{B}_j \prod_{j \text{ odd}} \hat{B}_j \prod_{j} \hat{A}_j + \mathcal{O}(\Delta t^2) \right)^n, \quad (2)$$

where $\hat{A}_j = e^{-i(-g\hat{\sigma}_j^x)\Delta t}$, and $\hat{B}_j = e^{-i(-J\hat{\sigma}_j^z\hat{\sigma}_{j+1}^z)\Delta t}$. $\Delta t$ is the Trotter step size given as $\Delta t = t/n$.

In a second-order symmetric Trotterization, the time evolution operator can be rewritten in [10]

$$e^{-i\hat{H}t} = \left(\prod_j \hat{A}_j \prod_{j\,\text{odd}} \hat{B}_j \prod_{j\,\text{even}} \hat{C}_j \prod_{j\,\text{odd}} \hat{B}_j \prod_j \hat{A}_j + \mathcal{O}(\Delta t^3)\right)^n, \quad (3)$$

where $\hat{A}_j = e^{-i(-g\hat{\sigma}_j^x)\frac{\Delta t}{2}}$, $\hat{B}_j = e^{-i(-J\hat{\sigma}_j^z\hat{\sigma}_{j+1}^z)\frac{\Delta t}{2}}$, and $\hat{C}_j = e^{-i(-J\hat{\sigma}_j^z\hat{\sigma}_{j+1}^z)\Delta t}$. The Trotter error for the symmetric Trotterization is of order $\mathcal{O}(\Delta t^3)$, which is expected to be less prominent than the first-order Trotter error of order $\mathcal{O}(\Delta t^2)$ [10,23,37]. Throughout this work, a five-spin chain ($N = 5$) is employed, and $|\downarrow\downarrow\downarrow\downarrow\downarrow\rangle$ is employed as the initial state $|\Psi(0)\rangle$. Once the spin system is prepared in the ground state in absence of transverse field, the transverse field is turned on; then, the system evolves in time with $\Delta t = 0.2/J$, and the Trotter errors are measured at later times to investigate the merit of the symmetric Tortterization.

## 3. Results and discussion

Figure 1 shows quantum circuits corresponding to the single Trotter steps for the first-order Trotterization and the symmetric Trotterization [10]. Trotterized DQSs are implemented via Qiskit [38], which provides the complete set of tools for interacting with superconducting quantum processors on IBM Quantum Experience [39]. Ideal quantum simulations with the Trotterizations are performed using *statevector_simulator*, which is a perfect quantum emulator excluding quantum errors, such as logical gate error, readout error, relaxation, and dephasing. Experimental quantum simulations with the Trotterizations are performed using *ibmq_santiago*, which is one of the IBM Quantum processors, with five superconducting qubits. The number of shots (how many times to run the circuit) is set to 1024 for statistics. As a reference, the exact, classical simulations are performed using QuSpin [40], in which the Trotter error is completely ruled out.

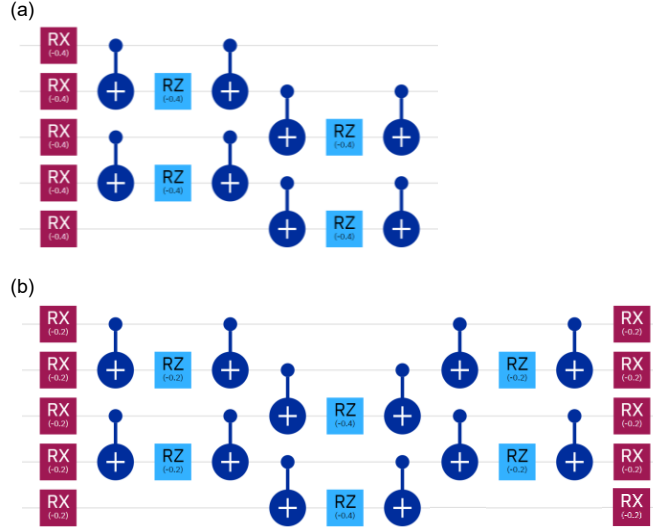

Figure 1. Quantum circuits for single Trotter steps of (a) a first-order Trotterization and (b) a second-order symmetric Trotterization ($N = 5$, $\Delta t = 0.2/J$, $g = J$). The circuits are composed of $R_x(\theta) = e^{-i\frac{\theta}{2}\hat{\sigma}_j^x}$ (rotation around X-axis), $R_z(\theta) = e^{-i\frac{\theta}{2}\hat{\sigma}_j^z}$ (rotation around Z-axis), and CNOT (controlled-NOT) gates.

To measure Trotter error in DQS of quantum dynamics of TFIM, we estimated total magnetization

$$M(t) = \left\langle \Psi(t) \left| \frac{1}{N} \sum_j \hat{\sigma}_j^z \right| \Psi(t) \right\rangle \quad (4)$$

and local magnetization for the $j$th spin

$$M_j(t) = \langle \Psi(t)|\hat{\sigma}_j^z|\Psi(t)\rangle. \tag{5}$$

The time evolution of the total magnetization for the first-order Trotterization and the symmetric Trotterization shows apparent deviation from the exact, classical simulation (Figure 2). The five spins evolve in time, starting from their initial state $|\downarrow\downarrow\downarrow\downarrow\downarrow\rangle$, corresponding to $M(0) = -1$ (see Figure 3 for the exact dynamics of each spin). In the presence of the transverse field, the spin state deviates from its initial state because the initial state is not an eigenstate anymore, and each spin oscillates. The spins oscillate rapidly when the transverse field $g$ is strong. Eventually, they will align with the transverse field, resulting in $M(\infty) = 0$, regardless of the strength of the transverse field. The deviation of the ideal results from the exact results originates from the Trotter error, irrelevant to the quantum errors caused by low fidelity. For $g = 1J$, Trotter error is negligible. As $g$ increases, the Trotter error becomes prominent, indicating that one needs a small Trotter step size to simulate rapidly fluctuating spins. Therefore, by modulating $g$, one can readily take into account different degrees of Trotter error. Compared with the ideal results, the experimental results exhibit more significant errors due to the quantum gate errors. However, it was not straightforward to differentiate the first-order Trotterization and the symmetric Trotterization in terms of the total magnetization.

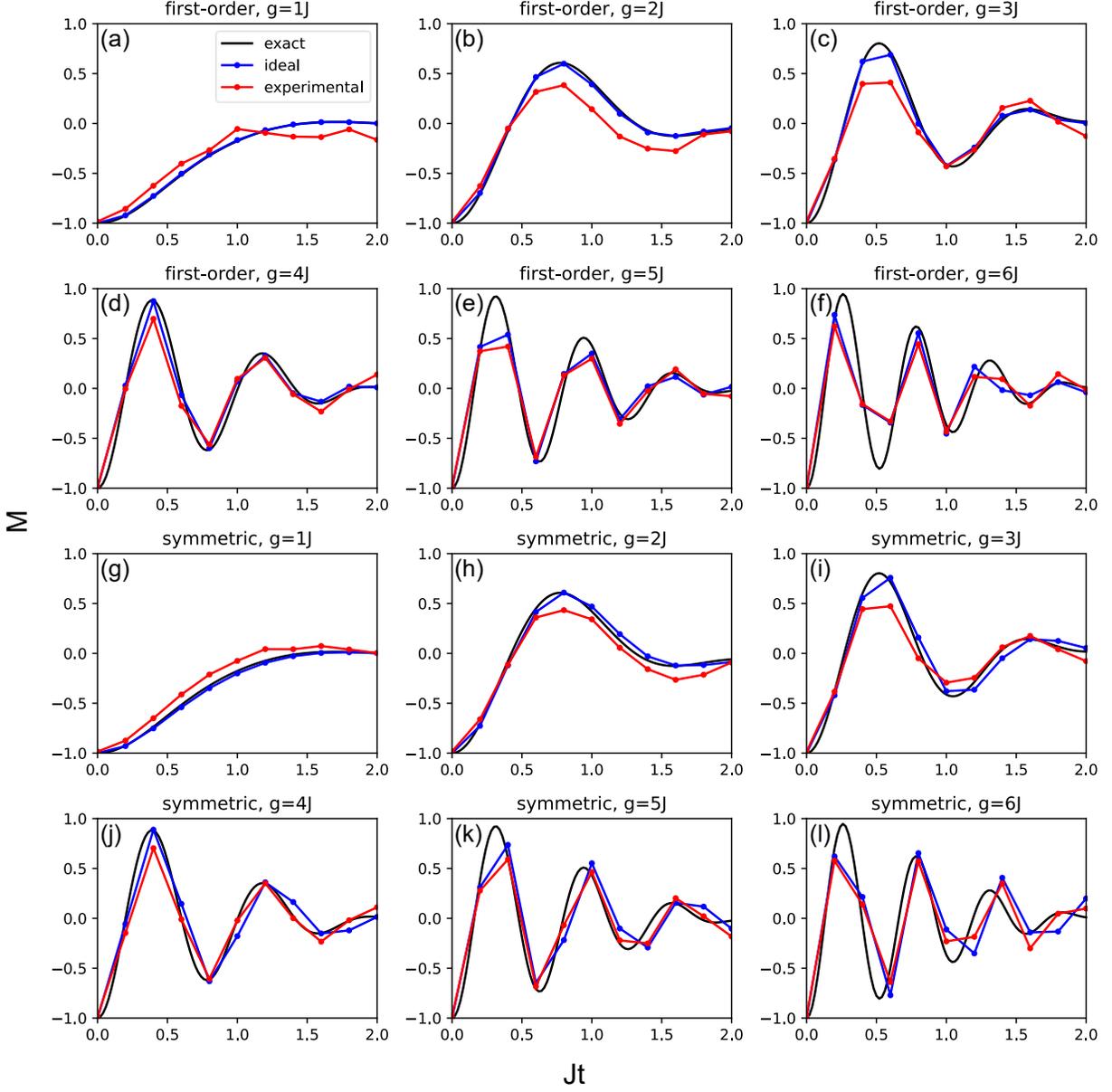

Figure 2. Total magnetization *M* vs. time for (a-f) the first-order Trotterization and (g-l) the symmetric Trotterization with different transverse fields *g* ranging from 1*J* to 6*J*. The exact dynamics is obtained in the exact, classical simulations. Ideal and experimental quantum simulations are performed using a perfect quantum emulator, *statevector_simulator*, and an IBM Quantum processor, *ibmq_santiago*, respectively.

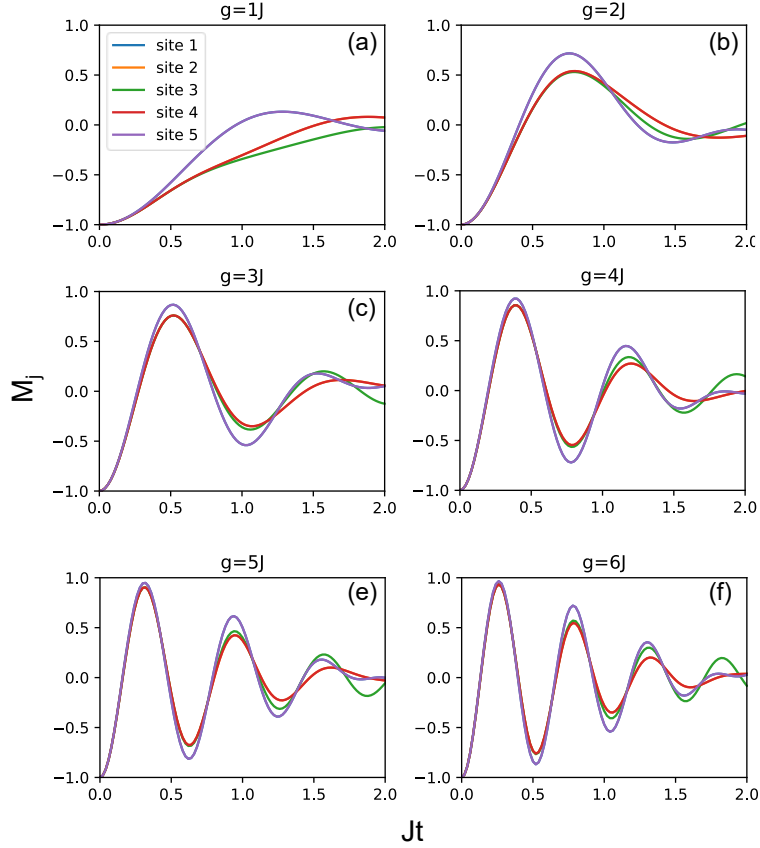

Figure 3. Local magnetization $M_j$ vs. time with different transverse fields $g$, resulting from the exact, classical simulations using QuSpin.

For further analysis of the difference between the first-order Trotterization and the symmetric Trotterization, local spin magnetizations $M_j$ were extracted from the results of DQS. Figure 4 displays time- and site-resolved errors of the ideal quantum simulation. The local magnetization plots show the Trotter error raising with the increase in $g$, more evidently than Figure 2. Interestingly, the symmetric Trotterization exhibits a more prominent Trotter error than the first-order Trotterization, and it is noticeable when the Trotter error is substantial with large $g$. Figure 5 shows time- and site-resolved errors of the experimental quantum simulation, resulting from quantum errors induced by gate operations on top of the Trotter error. In this case, the quantum errors seem to be predominant, compared with the Trotter error, especially for small $g$.

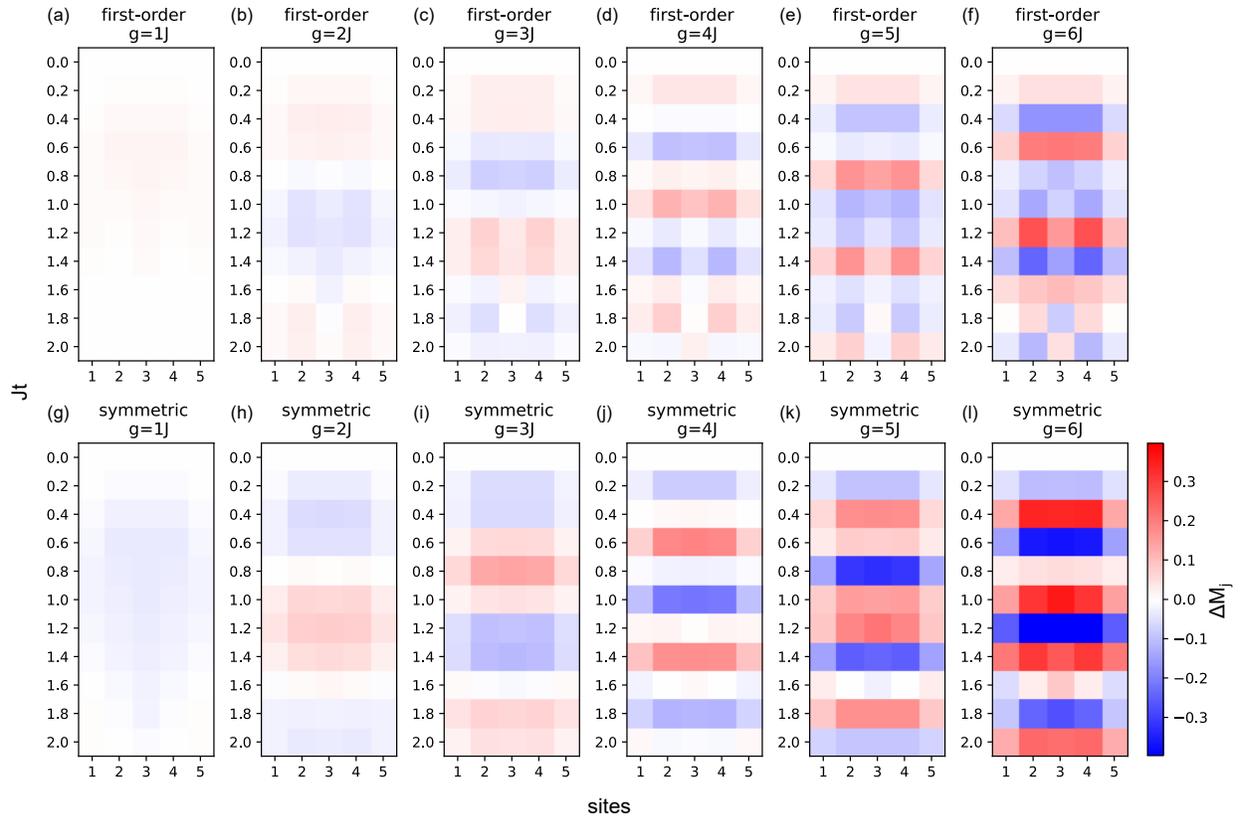

Figure 4. Time- and site-resolved errors for the local magnetization $M_j$, where the ideal quantum simulations are performed. The errors $\Delta M_j$ are deviations from the exact $M_j$.

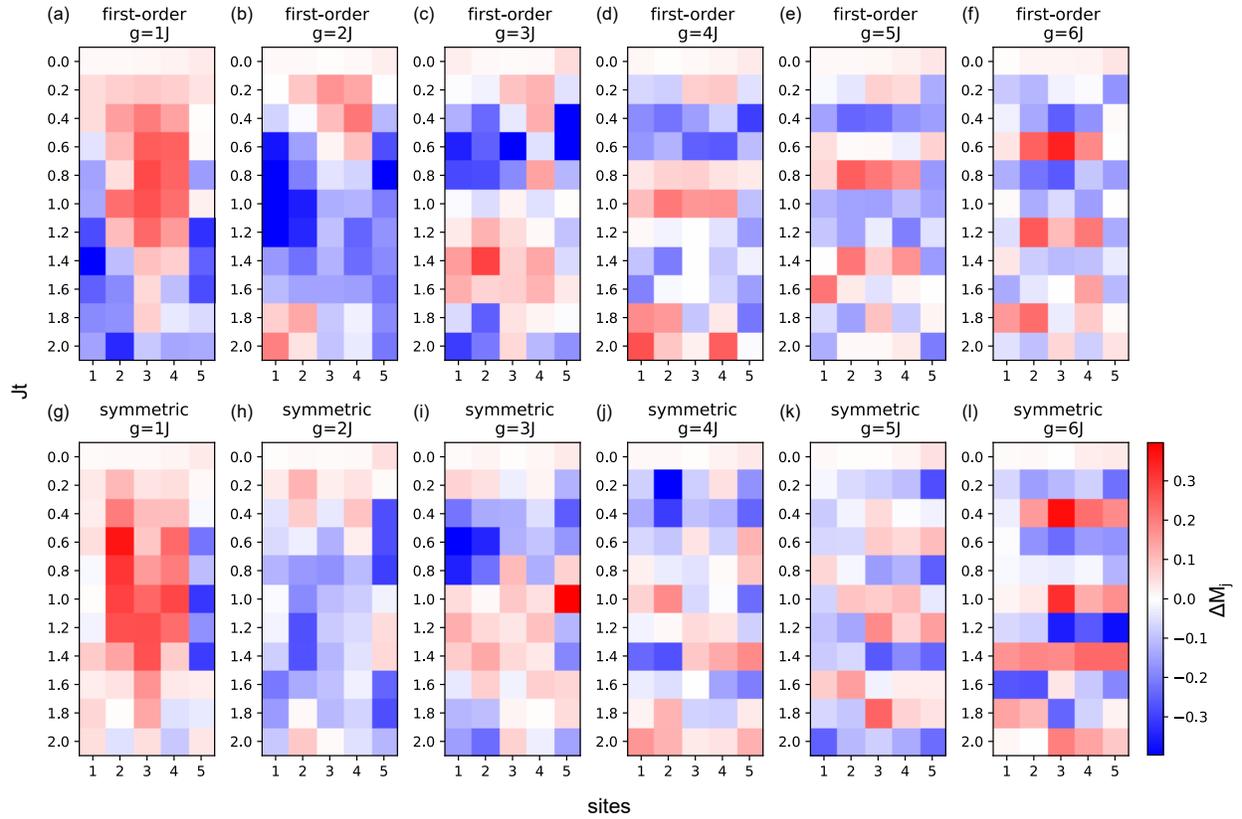

Figure 5. Time- and site-resolved errors for the local magnetization $M_j$, where the experimental quantum simulations are performed. The errors $\Delta M_j$ are deviations from the exact $M_j$.

Root mean square error (RMSE) is a useful metric to quantitatively analyze errors. Figure 6a shows RMSE based on the errors in the local magnetization shown in Figure 4 and Figure 5 (see Figure 6b for RMSEs based on the errors in the total magnetization). The ideal quantum simulation results present that the RMSE increases monotonically as $g$ increases, as already demonstrated in the time- and site-resolved plots. It is also shown clearly that the RMSE of the symmetric Trotterization is almost twice larger than that of the first-order Trotterization. Although it is expected that a higher-order Trotterization mitigates the Trotter error, it turns out not to be the case for quantum dynamics simulation of transverse-field Ising chains. Since the Trotterized circuits employed in this work were not ordered in the optimal way [24,37], there would be room for improvement for the symmetric Trotterization. Thus, further systematic and analytical studies in consideration of the optimization can follow to investigate high-order Trotterizations for various quantum circuits comprehensively. Notwithstanding, a major lesson that we learn from the current results is that one should carefully determine whether it is worthwhile to adopt a higher-order Trotterization.

In the experimental quantum simulation, the $g$ dependence and Trotterization-scheme dependence are ruled out because of prominent quantum gate errors of the employed superconducting quantum computer (Figure 6). Even at the very beginning of the dynamics, where the circuit depth is still low, we observe a significant contribution of the quantum errors (Figure 5). This is an indication of the capability and limitation of current quantum devices at an early stage of NISQ, leading to the vital need for quantum error mitigation techniques [10,12,41].

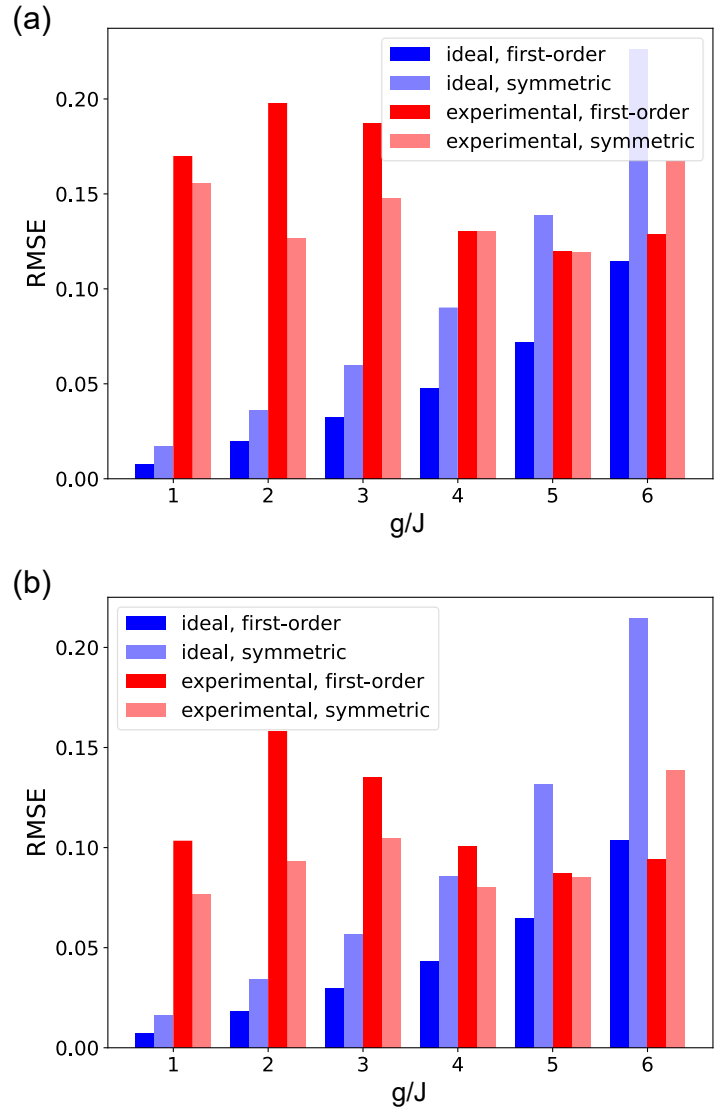

Figure 6. RMSEs with different $g$, extracted from the ideal and experimental quantum simulations with the first-order and symmetric Trotterizations. (a) The RMSEs are computed with the local spin magnetization. (b) The RMSEs are based on the errors in the total magnetization shown in Figure 2.

## 4. Conclusion

In summary, quantum simulations were performed with the second-order symmetric Trotterization to investigate the effectiveness of a high-order Trotterization for DQS of the quantum many-body spin dynamics of TFIM. Although the second-order symmetric Trotterization is expected to mitigate the Trotter error compared with the first-order Trotterization, it turned out that the symmetric Trotterization does not reduce the Trotter error in the DQS of the quantum dynamics of TFIM in the given conditions. There could be room for improvement by optimizing the Suzuki-Trotter decomposition. However, in the absence of optimization, the symmetric Trotterization can end up with more erroneous results than the first-order Trotterization in particular applications, regardless of the quantum errors. Apart from that, the effect of the quantum errors on the experimental quantum simulations was investigated using an actual superconducting quantum processor. The experimental quantum simulations showed that the circuits for the first-order Trotterization and the symmetric Trotterization brought about a similar magnitude of error because the primary sources of error were the quantum gate operations rather than the Trotter error, implying that the Trotter error would play a minor role in infancy on NISQ devices. At the end of the day, while presenting a pedagogical way to perform quantum simulation using a real quantum computer, this work disclosed that a higher-order Trotterization should be carefully tested and employed to take advantage of higher-order Trotterizations in tandem with near-term NISQ devices exhibiting decently-suppressed quantum errors.

**Acknowledgment**

This work was supported by Incheon National University Research Grant in 2021. I acknowledge use of the IBM Quantum Experience for this work. The views expressed are those of the author and do not reflect the official policy or position of IBM or the IBM Quantum Experience team.